\documentclass[11pt,graphicx]{article}
\usepackage{amssymb,amsmath,amsfonts}
\usepackage{graphicx}
\usepackage{graphics}
\usepackage{eepic,epsfig}
\begin{document}
\title{Spherically symmetric vacuum solutions of modified gravity theory in higher dimensions}
\author{T. R. P. Caram\^es {\thanks{E-mail: carames@fisica.ufpb.br }}  
and E. R. Bezerra de Mello \thanks{E-mail: emello@fisica.ufpb.br}\\
Departamento de F\'{\i}sica-CCEN, Universidade Federal da Para\'{\i}ba\\
58.059-970, C. Postal 5.008, J. Pessoa, PB,  Brazil}
\maketitle
\begin{abstract}
In this paper we investigate spherically symmetric vacuum solutions of $f(R)$ gravity in a higher dimensional spacetime. With this objective we construct a system of non-linear differential equations, whose solutions depend on the explicit form assumed for the function $F(R)=\frac{df(R)}{dR}$. We explicit show that for specific classes of this function exact solutions from the field equations are obtained; also we find approximated results for the metric tensor for more general cases admitting $F(R)$ close to the unity.  
\vspace{1pc}
\end{abstract}
\maketitle
\section{Introduction}
One of most fascinating problem in modern Cosmology is to explain satisfactorily the accelerated expansion of the Universe \cite{Ries}. In this way three main candidates have been considered in literature: The presence of a cosmological constant $\Lambda$, the existence of a dark energy and the modified theories of gravity. The cosmological constant variant brings naturally the cosmological problem and the coincidence problem as well. Requiring that $\Lambda$ presents a tiny value to cause the present acceleration demands to an extreme fine-tuning. In the second scenario, it is postulated the existence of a fluid with equation of state $P\approx -\rho$, being $\rho$ and $P$ the energy density and pressure, respectively, which comes to dominate late in the matter era. The third alternative opens the possibility to provide new contributions to the equation of state compatible with an accelerated expansion of the Universe. The main idea of modified gravity theories resides in the generalization of the Einstein-Hilbert action in large scale, admitting a scalar non-linear function of the Ricci scalar curvature, $f(R)$.\footnote{The idea of modifying Einstein-Hilbert action has already been proposed many years ago \cite{Ste}, as a possible approach to construct a renormalized theory of gravity.} The modified gravity also allows a unification of the early-time inflation \cite{Staro} and late-time cosmic speed-up \cite{Carrol,Fay}. \footnote{Specifically in \cite{NO,NO1}, the authors proposed realistic models which support the inflationary epoch with the present cosmic acceleration.} These models seem also relevant to explain the hierarchy problem and unification of Grand Unified Theories, GUTs, with gravity \cite{Cognola}. Such theories avoid the Ostrogradski's instability \cite{Ostr} that can otherwise prove to be problematic for general higher derivatives theories \cite{Woo}. 

In the context of modified theories of gravity, static spherically symmetric solutions of the field equations have been analysed in vacuum sector \cite{Mut} as well as in the presence of perfect fluid source \cite{Mut1}, respectively; in the later, the authors show that a given distribution of matter is not capable to uniquely determine $f(R)$. In \cite{Capo}, spherically symmetric solutions of $f(R)$ theories of gravity have been analysed by using the Noether symmetry. Moreover, static cylindrically symmetric vacuum solution has been analysed in \cite{Azad}.

Another attractive model to describe our Universe is the braneworld scenario. By this model our world is represented by a four-dimensional sub-manifold, a three-brane, embedded in a higher dimensional spacetime \cite{Akama,Rubakov}. In fact the conjecture that our Universe may have more than four dimensions was proposed by Kaluza \cite{Kaluza} with the objective to unify  gauge theories with gravitation in a geometric formalism. In this context extra coordinates are compactified to circles of small radius that would be observable at high-energy scale comparable to the Planck one. Many of high energy theories of fundamental physics are formulated in higher-dimensional spacetime. In supergravity and superstring the idea of extra dimensions has been extensively used. Also in higher-dimensional spacetime context, of particular interest are the models introduced by Randall and Sundrum \cite{RS,RS1}. In these models it is assumed that all matter fields are confined on the brane and gravity only propagates in the five dimensional bulk. In the RSI model, the hierarchy problem between the Planck scale and the electroweak one is solved if the distance between the two branes is about $37$ times the AdS radius.

In this paper we join both promising ideas about the Universe. In this sense we shall analyse the modified theories of gravity in a higher dimensional spacetime under classical point of view. Specifically, as the first starting point, we shall analyse vacuum solutions admitting spherically symmetric {\it Ansatz} for the metric tensor. In this sense our approach is a higher dimensional extension of that one presented in \cite{Mut}. We shall adopt radial functions for the function $F(R)$ and calculate the corresponding results for the metric tensor. 

This paper is organized as follows. In Section \ref{sec2} we present for a $(1+d)-$dimensional spacetime, spherically symmetric fields equations for general modified theories of gravity. We also present some relevant properties related with the geometry of the corresponding spacetime. In Section \ref{sec3} we provide explicitly vacuum solutions of the field equations for a large class of radial function $F(R)$. We also analyse the specific case with $d=2$. We show that differently from Einstein equation, non-trivial vacuum solutions are obtained for the latter. Moreover; comparing this case with large value of $d$ ones, we show that there exist a substantial modification on the field equations, which justify a separated analysis. Finally we summarize our most important results in Section \ref{sec4}. The Appendix \ref{A} contains some technical details useful in Section \ref{sec3}. In this paper we use signature $+2$, and the definitions: $R_{\beta\gamma\delta}^\alpha=\partial_\gamma\Gamma_{\beta\delta}^\alpha-...$, $R_{\alpha\beta}= R_{\alpha\gamma\beta}^\gamma$. We also use units with $G=c=1$.

\section{Field equations in higher-dimensional spherically symmetric spacetime}
\label{sec2}
In this section we shall analyse the modified gravity action in a higher-dimensional spacetime. Specifically static spherically symmetric metric tensor will be considered. We shall present some geometric properties associated with this corresponding spacetime, and provide the general fields equations obeyed by the components of the metric tensor.
 
Let us consider first the action below for modified theories of gravity:
\begin{equation}
\label{Act}
S=\frac1{2\kappa}\int d^{d+1}x\sqrt{-g}f(R)+{\cal S}_m \ ,
\end{equation} 
where $R$ is the Ricci scalar, $\kappa=8\pi G$, and ${\cal S}_m$ represents the action associated with matter fields. By using the metric formalism, the field equations becomes:
\begin{equation}
\label{G}
G_{\mu\nu}\equiv R_{\mu\nu}-\frac{1}{2}Rg_{\mu\nu}=T^{c}_{\mu\nu}+\kappa T^{m}_{\mu\nu}	\ ,
\end{equation}
in which $T^{c}_{\mu\nu}$ is the geometric energy-momentum tensor, namely
\begin{equation}
\label{3}
T^{c}_{\mu\nu}=\frac{1}{F(R)}\left\{\frac{1}{2}g_{\mu\nu}\left(f(R)-F(R)R\right)+\nabla^{\alpha} \nabla^{\beta} F(R)\left(g_{\alpha\mu}g_{\beta\nu}-g_{\mu\nu}g_{\alpha\beta}\right)\right\}      
\end{equation}
with $F(R)\equiv\frac{df(R)}{dR}$. 

The standard minimally coupled energy-momentum tensor $\tilde{T}^{m}_{\mu\nu}$, derived from the matter action, is related to $T^{m}_{\mu\nu}$ by
\begin{equation}
\label{T}
T^{m}_{\mu\nu}=\tilde{T}^{m}_{\mu\nu}/F(R) \ .
\end{equation}

In absence of matter fields the field equations reads
\begin{equation}
\label{FE}
F(R)R_{\mu\nu}-\frac{1}{2}f(R)g_{\mu\nu}-\nabla_{\mu}\nabla_{\nu}F(R)+g_{\mu\nu}\Box F(R)=0 \ . 
\end{equation}
Taking the trace of the above equation we get 
\begin{equation}
\label{6}
F(R)R-\frac{(d+1)}{2}f(R)+d\Box{F(R)}=0 \ ,
\end{equation}
which express a further degree of freedom that arises in the modified theory, namely the scalar curvature one. Through this equation it is possible to express $f(R)$ in term of its derivatives, as follows
\begin{equation}
\label{fr}
f(R)=\frac{2}{d+1}(F(R)R+d\Box{F(R)}) \ .
\end{equation}

Substituting the above expression into (\ref{FE}) we obtain
\begin{align}
	F(R)R_{\mu\nu}-\nabla_{\mu}\nabla_{\nu}F(R)=\frac{g_{\mu\nu}}{d+1}\left[F(R)R-\Box F(R)\right] \ .
\end{align}
From this expression we can see that the combination below
\begin{align}
\label{A0}
	A_\mu=\frac{F(R)R_{\mu\mu}-\nabla_{\mu}\nabla_{\mu}F(R)}{g_{\mu\mu}} \ ,
\end{align}
with fixed indices, is independent of the corresponding index. 

\subsection{Spherically symmetric metric tensor}
Adopting the coordinate system $x^\mu=(t, r, \theta_1, \theta_2, ...  ,\theta_{d-2}, \phi)$, with $d\geq3$ and the coordinates defined in the intervals $t\in(-\infty, \infty)$, $\theta_i\in [0, \pi]$ for $i=1, 2 ... d-2$, $\phi\in[0, 2\pi]$ and $r\geq0$, the static metric tensor associated with the most general spherically symmetric $(1+d)-$dimensional spacetime, can be represented by  
\begin{eqnarray}
\label{Metric}
g_{00}&=&-u(r) \nonumber \\ 
g_{11}&=&v(r) \nonumber \\ 
g_{22}&=&r^2 \nonumber\\ 
g_{jj}&=&r^2\sin^2\theta_1\sin^2\theta_2 ...\sin^2\theta_{j-2} \ ,
\end{eqnarray}
for $3\leq j \leq d$, and $g_{\mu\nu}=0$ for $\mu\neq \nu$. For this metric tensor the non-vanishing components of the Ricci tensor read:
\begin{equation}
\label{00}
R_0^0=\frac{1}{2}\left[\frac{(u')^{2}}{2u^2v}+\frac{u'v'}{2uv^{2}}-\frac{u''}{uv}-\frac{(d-1)u'}{uvr}\right] \ ,
\end{equation}
\begin{equation}
\label{11}
R_1^1=\frac{1}{2}\left[\frac{u'v'}{2uv^2}+\frac{(d-1)v'}{rv^2}-\frac{u''}{uv}+\frac{(u')^{2}}{2u^{2}v}\right] \ ,
\end{equation}
\begin{equation}
\label{22}
R_2^2=\frac{v'}{2rv^{2}}-\frac{u'}{2ruv}-\frac{(d-2)}{r^2v}+\frac{(d-2)}{r^2} \ .
\end{equation} 
The spherical symmetry requires that
\begin{equation}
R^{2}_{2}=R^{3}_{3}=R^{4}_{4}=...=R^{d}_{d} \ ,
\end{equation}
which implies that the scalar curvature will be given by
\begin{eqnarray}
\label{Ricci}	
R&=&-\frac{2}{r^{2}v}\left\{\frac{d(d-3)+2}{2}-\frac{v\left[d(d-3)+2\right]}{2}+\frac{(d-1)}{2}\left(\frac{u'}{u}-\frac{v'}{v}\right)r\right. \nonumber\\
&-&\left.\frac{1}{4}\frac{u'}{u}r^{2}\left(\frac{u'}{u}+\frac{v'}{v}\right)+\frac{1}{2}r^{2}\frac{u''}{u}\right\} \ .
\end{eqnarray} 
In all the above equations the primes corresponds derivative of the function with respect to the radial coordinate.

Defining $Y(r)=u(r)v(r)$ and by using the identity $A_\mu=A_\nu$ for all $\mu$ and $\nu$, we can construct the two linearly independent homogeneous differential equations below:
\begin{equation}
\label{Y1}
	2rF^{''}-r\frac{Y'}{Y}F'-(d-1)\frac{Y'}{Y}F=0 \ ,
\end{equation}
and
\begin{eqnarray}
\label{Y2}
	&-&4(d-2)u+4Y(d-2)-4ru\frac{F'}{F}+2r^{2}u'\frac{F'}{F}-r^{2}u'\frac{Y'}{Y} \nonumber\\
	&+&2r^{2}u^{''}+2ru'(d-3)+2ur\frac{Y'}{Y}=0 \ .
\end{eqnarray}	 
These equation will be the basis for out future analysis.
\section{Vacuum solutions}
\label{sec3}
In this section we shall analyse possible vacuum solutions of the field equations with radial symmetry. As we shall see, these solutions strongly dependent on the explicit form adopted for $F(R)$. 

\subsection{Spacetimes with constant scalar curvature}
As first part of our investigation, let us consider spacetimes with constant scalar curvature. For this case, the field equations (\ref{Y1}) and (\ref{Y2}) can be written as 
\begin{equation}
\label{eq1}
uv'+vu'=0	
\end{equation}
and
\begin{eqnarray}
\label{eq2}
(1-v)(d-2)+\frac{r}{2}\left(\frac{u'}{u}+\frac{v'}{v}\right)\left(\frac{r}{2}\frac{u'}{u}-1\right)-\frac{1}{2}\frac{r^{2}u''}{u}-
\frac{ru'(d-3)}{2u}=0 \ ,
\end{eqnarray}
whose solutions are shown to be given by
\begin{equation}
\label{eq3}
v(r)=\frac{c_{0}}{u(r)}	\ \ {\rm with} \ \ u(r)=c_{0}+\frac{c_{1}}{r^{d-2}}+c_{2}r^{2} \ ,
\end{equation}
where $c_{0}$, $c_{1}$ and $c_{2}$ are integration constants. In order to maintain unchanged the signature we impose that $c_{0}>0$. By using (\ref{Ricci}), we can find that the corresponding scalar curvature is $R=-d(d+1)c_{2}/c_{0}$. It is possible to rescale the time coordinate so that we can always choose $c_{0}=1$. A Lagrangian in which $f(R)=R+\Lambda$ gives rise to Schwarzschild-de Sitter (SdS) solution, that means a Schwarzschild spacetime in the presence of a cosmological constant. 

For a $d-$dimensional space, the Newtonian potential obeys the equation
\begin{equation}
\label{Pot}
\nabla^{2}\Phi=\Omega_dM\delta^{(d)}(\stackrel{\rightarrow}{r}) \ ,
\end{equation}
being $M$ the mass of the particle placed at origin. However, in the case where $d\geq3$ it is known that\footnote{The $d=2$ case, will be analysed separately in subsection \ref{d2}.}
\begin{equation}
\nabla^{2}\frac{1}{r^{d-2}}=-(d-2)\Omega_{(d)}\delta^{(d)}(\stackrel{\rightarrow}{r}) \ ,
\end{equation}
where $\Omega_{(d)}=\frac{2\pi^{d/2}}{\Gamma(\frac{d}{2})}$. So, by using the above relations this potential is written as
\begin{equation}
\label{Pot1}
\Phi=-\frac{M}{(d-2)}\frac{1}{r^{d-2}} \ .
\end{equation}

Now coming back to the SdS-type solution, we have the forms:
\begin{eqnarray}
v(r)=\frac1{u(r)} \ \ {\rm with} \ \ u(r)=1-\frac{2M}{(d-2)}\frac1{r^{d-2}}-\frac{\Lambda r^2}{d(d-2)} \ ,
\end{eqnarray}
whose corresponding scalar curvature is
\begin{equation}
R=2\Lambda\frac{(d+1)}{(d-1)} \ .
\end{equation}
This solution is achieved in (\ref{eq3}) by choosing $c_{1}=-2M/(d-2)$ and $c_{2}=-2\Lambda/d(d-1)$. Because we are deal with spherically symmetric solutions, the idea of horizon associated with the corresponding black hole, takes place. These regions are found by imposing the condition $u(r)=0$; however for a higher dimensional spacetime this condition leads us to high order algebraic equations whose solutions, in general, are not provided by any analytical procedure. 
 
\subsection{Solutions with $u(r)v(r)$=constant}
Another class of solutions can be obtained by imposing $Y(r)=u(r)v(r)=Y_{0}=$const. In this case a solution of  (\ref{Y1}) is $F(r)=Ar+B$.\footnote{In this paper we shall use $F(r)$ to represent the radial function obtained from $F(R(r))$.} Substituting this expression into (\ref{Y2}), we obtain an exact solution for $u(r)$ given by\footnote{The details of this calculation are in Appendix \ref{A}.}
\begin{eqnarray}
\label{14}
u(r)&=&Y_{0}+\frac{1}{2}\frac{A}{B^{2}}c_{1}-\frac{A}{B^{3}}\left(\frac{2B^{2}Y_{0}}{(d-1)}+Ac_{1}\right)r+\frac{(-1)^{d}}{dBr^{d-2}}\left(\frac{A}{B}\right)^{3-d}c_{1}+c_{2}r^{2}\nonumber\\
&+&\frac{c_{1}}{B}\sum^{d-1}_{n=3}\frac{(-1)^{n}}{nr^{n-2}}\left(\frac{A}{B}\right)^{3-n}+r^{2}\frac{A^{2}}{B^{4}}\left(\frac{2B^{2}Y_{0}}{(d-1)}+Ac_{1}\right)\ln\left(1+\frac{B}{Ar}\right)\ ,
\end{eqnarray}
where $c_{i}$ are integrations constants. In order to have a SdS-type solutions for the non-trivial gravity theory, the Newtonian potential-type term requires that $c_{1}=-\frac{2dMB}{(-1)^{d}(d-2)}\left(\frac{A}{B}\right)^{d-3}$. We can define the independent term $Y_{0}+\frac{1}{2}\frac{A}{B^{2}}=1$, which leads $Y_{0}\neq 1$. In this case, we shall have a correction in the Newtonian potential due the term $Ac_{1}/2B^{2}\neq 0$. It is natural to choose $c_{1}=-\frac{2dMB}{(-1)^{d}(d-2)}\left(\frac{A}{B}\right)^{d-3}$ in order the form of Newtonian potential be achieved. In this case we have:
\begin{equation} 
\label{15}
Y_{0}=\frac{d-1}{d-2}
\end{equation}
and  
\begin{equation}
\label{16}
A=\left(\frac{(-1)^{d}}{dM}\right)^{\frac{1}{(d-2)}} \ ,
\end{equation}
where we have set $B=1$. This leads to the following form for $F(r)$:
\begin{equation}
\label{17}
F(r)=1+\left(\frac{(-1)^{d}}{dM}\right)^{\frac{1}{(d-2)}}r \ .
\end{equation}
Finally, (\ref{14}) will be given by
\begin{eqnarray}
\label{18}
u(r)=1-\frac{2M}{(d-2)r^{d-2}}-\frac{2\Lambda}{d(d-1)}r^{2}
	+\frac{2dM}{d-2}\sum^{d-1}_{n=3}\frac{(-1)^{n+1-d}}{nr^{n-2}}\left(\frac{(-1)^d}{dM}\right)^\frac{d-n}{d-2} \ .
\end{eqnarray}
Since $Y(r)=u(r)v(r)$, through (\ref{15}) we can straightforwardly determine $v(r)$. The corresponding scalar curvature is given by:
\begin{equation}
\label{19}
R=\frac{d-2}{r^{2}}+\frac{2\Lambda (d-2)(d+1)}{(d-1)^{2}}+\sum^{d-1}_{n=3}\frac{M^{\frac{n-2}{d-2}}c_{n,d}}{r^{n}} \ ,
\end{equation}
which depends on the cosmological constant and the mass parameter. Moreover, one can obtain $f(R)$ by using (\ref{Y1}). So, the exact form, in terms of the radial coordinate, $r$, for $f(r)$ will be given by
\begin{eqnarray}
\label{radial}
f(r)&=&\frac{2}{(d+1)}\left[\frac{d-2}{r^2}+\frac{2\Lambda(d-2)(d+1)}{(d-1)^2}+\left(\frac{(-1)^{d}}{dM}\right)^{\frac{1}{(d-2)}}\frac{(d-2)(d+1)}{r}\right.\nonumber\\
&-&\left.\frac{2(-1)^{\frac{d}{d-2}}}{(d-1)r^{d-1}}(Md)^{\frac{d-3}{d-2}}+\sum^{d-1}_{n=3}\left(\frac{M^{\frac{n-2}{d-2}}c_{n,d}}{r^n}+\frac{M^{\frac{n-3}{d-2}}p_{n,d}}{r^{n-1}}+\frac{M^{\frac{n-3}{d-2}}s_{n,d}}{r^{n-1}}\right)\right]\ .
\end{eqnarray} 
Where we have defined
\begin{equation}
\label{coef1}
c_{n,d}\equiv \frac{2d^{\frac{n-2}{d-2}}}{n(d-1)}(-1)^{3d-2(n+1)}\left\{\left[2(n-1)-d\right](d-1)-(n-2)(n-1)\right\}\ ,
\end{equation}
\begin{equation}
\label{coef2}
p_{n,d}\equiv\frac{2d^{\frac{n-3}{d-2}}}{n(d-1)}(-1)^{\frac{3d^2-d(2n+7)+4(n+1)}{d-2}}\left\{\left[2(n-1)-d\right](d-1)-(n-2)(n-1)\right\}\ ,
\end{equation}
and
\begin{equation}
\label{coef3}
s_{n,d}\equiv \frac{2d^{\frac{d+n-5}{d-2}}}{n(d-1)}(-1)^{\frac{2(2d-n-1)}{d-2}}(d-n+1)\ .
\end{equation}
It is important to observe that the coefficients $p_{n,d}, \ s_{n,d}$ and $c_{n,d}$ will be present only when $d\geq 4$, so $p_{n,3}=s_{n,3}=c_{n,3}=0$, $\forall n$. The expressions (\ref{19}) and (\ref{radial}) reproduce previous results when we substitute $d=3$. However,  in principle, for $d>3$, it is not a trivial matter to provide an explicit expression for $f(R)$ by using the above results. So, as an application of this formalism, let us consider $d=4$. For this case, we have the following exact results as shown below:
\begin{equation}
\label{20}
R=\frac{2}{r^2}+\frac{20\Lambda}{9}-\frac{8\sqrt{M}}{9r^3}
\end{equation} 
and
\begin{equation}
\label{21}
f(r)=\frac{4}{3r^2}+\frac{8\Lambda}{9}+\frac{2}{\sqrt{M}r}-\frac{8\sqrt{M}}{9r^3} \ .
\end{equation}   
It is possible, by using (\ref{20}) to express the radial coordinate, $r$, in terms of the scalar curvature, $R$; this provides a third order algebraic equation, which by its turn possesses one real root. Because the result found for this root is a very long one, the corresponding expression for $f(R)$ does not clarify the physical content of the system under investigation, in this way we shall only analyse the above expressions in two asymptotic regimes; they are: (i) for $r\gg\sqrt{M}$ and (ii) $r\ll\sqrt{M}$.\footnote{For $d=4$ the geometric mass has unity of length square.} Let us start with the first approximation. For this case the leading term in (\ref{20}) reads:
\begin{equation}
\label{scalar}
R\approx \frac{2}{r^2}+\frac{20\Lambda}{9} \ ,
\end{equation}
consequently
\begin{equation}
\label{faprox}
f(R)\approx \pm\frac{2}{3}\sqrt{\frac{9R-20\Lambda}{2M}}+\frac{8\Lambda}{9} \ .
\end{equation}
On the other hand, the second approximation leads:
\begin{equation}
\label{scalar}
R\approx -\frac{8\sqrt{M}}{9r^3}+\frac{20\Lambda}{9}
\end{equation}
and
\begin{equation}
\label{faprox1}
f(R)\approx R-\frac{4\Lambda}{3} \ .
\end{equation}
Finally we would like to say that, in principle, similar procedures as we have done before can also be applied for higher-dimensional spacetimes. 
\subsection{Different {\it Ans\"atze} for $F(r)$}
Although the most interesting models for modified gravity theories are those that reproduce Einstein theory of gravity for small disance, in this subsection we shall consider specific {\it Ans\"atze} for $F(r)$ that allow us to obtain exact solutions for the metric tensor.

\subsubsection{Solutions with $F(r)=F_{0}r$}
Considering $F(r)=F_{0}r$, Eq. (\ref{Y2}) provides an exact solution for $u(r)$. It is:
\begin{equation}
\label{22}
u(r)=Y_{0}\frac{(d-2)}{(d-1)}+\frac{c_{1}}{r^{d-1}}+c_{2}r^{2} \ .
\end{equation}
It would be interesting if we included in this analysis only purely de Sitter-type solutions. Pursuing this objective we may set $c_{1}=0$, $Y_{0}=\frac{(d-1)}{(d-2)}$ and $c_{2}=-\frac{2\Lambda}{d(d-1)}$. This provides
\begin{equation}
\label{23}
u(r)=1-\frac{2\Lambda}{d(d-1)}r^{2} \ .
\end{equation}
The scalar curvature associated with this solution is
\begin{equation}
\label{24}
R=\frac{(d+1)}{(d-1)}\Lambda+\frac{2+d(d-3)}{2r^{2}} \ ,
\end{equation}   
which leads, by means of the same procedure adopted in the previous section, to the following exact form for $f(R)$   
\begin{eqnarray}
\label{25}
f(R)&=&\pm \frac{2F_{0}\Lambda(d-3)}{(d-1)^2}\sqrt{\frac{(2+d(d-3))(d-1)}{2R(d-1)-2\Lambda(d+1)}} \nonumber\\ &\pm& \frac{\left[d(d-3)+2d(d-2)+2\right]F_0}{d+1}\sqrt{\frac{2(d-1)R-2(d+1)\Lambda}{(2+d(d-3))(d-1)}} \ .
\end{eqnarray}   
The corresponding result matches with that one obtained by T. Multamaki and I. Vilja in \cite{Mut} for $d=3$.      
\subsubsection{Solutions with $F(r)=F_{0}r^{n}$}   
    
Another {\it Ansatz} that we can investigate is that one in which $F(r)=F_{0}r^{n}$, for $n\neq 1$, whose corresponding solution for $u(r)$ is given by
\begin{equation}
\label{26}
u(r)=u_{0}r^{m} \ ,
\end{equation}
where $m\equiv \frac{2n(n-1)}{n+d-1}$; furthermore we have
\begin{equation}
\label{27}
Y(r)=\frac{(d^2+d(2n-3)+2(n-1)^{2})(d-(n-1)^{2})}{(d-2)(n+d-1)^{2}}u_{0}r^{m} \ .
\end{equation}
With the help of the equations (\ref{26}) and (\ref{27}) we can compute the scalar curvature. It is: 
\begin{equation}
\label{28}
R=\frac{dn(d-2)(n-2)}{n(n-2)-(d-1)}\frac{1}{r^{2}} \ ,
\end{equation}
from which we can obtain the corresponding $f(R)$ given by
\begin{equation}
\label{29}
f(R)=\frac{2F_{0}{h}_{n,d}}{(2-n)}\left(\frac{dn(d-2)(n-2)}{n(n-2)-(d-1)}\right)^{\frac{n}{2}}R^{1-\frac{n}{2}} \ .
\end{equation}
Where the constant $h_{n,d}$ is defined by
\begin{equation}
\label{coef4}
{h}_{n,d}\equiv\frac{2d^2(n-1)+6n(n-1)+d(d^2-1)-2(n^3-1)}{(d+1)(d(d-3))+2n(d+n)-2(2n-1)} \ .
\end{equation}
However, in an empty space this constant has no physical relevance, because it can be dropped out when the action is extremized. 

\subsubsection{Solutions with $F(r)=1+Ar^{n}$ with $Ar^{n}< 1$}
As we have already mentioned, in general it is a difficult matter to obtain an explicit form for $f(R)$; the reason is because some expressions found for the scalar curvature do not allow to express the radial coordinate, $r$, algebraically in terms of $R$. This is what happens in the present analysis when we use $F(r)=1+Ar^{n}$. A possible way to circumvent this problem, is to adopt an approximated procedure. In this direction we shall consider that the system under consideration is a small correction on General Relativity Theory (GR); then we assume $F(r)=1+Ar^{n}$ with $Ar^{n}< 1$. In this case, we were able to find an approximated solution for $u(r)$:
\begin{equation}
\label{44}
u(r)=1-\frac{2M}{(d-2)}\frac{1}{r^{d-2}}-\frac{2\Lambda}{d(d-1)}r^{2}+\frac{2Ar^n}{(d-1)}\left(\frac{Md}{(d-2)r^{d-2}}-1\right)+O((Ar^n)^2) \ .
\end{equation}
The expression for $Y(r)$ can be obtained from $F(r)$ by using (\ref{Y1}). In this case, taking into account the adopted approximation, we have:
\begin{equation}
\label{45}
Y(r)=1+\frac{2(n-1)}{d-1}Ar^{n}+O((Ar^n)^2) \ ,
\end{equation}
consequently, the scalar curvature can also be found approximately by
\begin{equation} 
\label{46}
R=\frac{2\Lambda (d+1)}{d-1}+Ar^{n}\left(B^{(0)}_{n,d}+\frac{B^{(2)}_{n,d}}{r^{2}}+\frac{B^{(d)}_{n,d}}{r^{d}}\right)+O((Ar^n)^2) \ .
\end{equation}
In the above expression we have defined $B^{(0)}_{n,d}=\frac{2dn(n+d-2)}{(d-1)}$, $B^{(2)}_{n,d}=-\frac{4\Lambda \left(d(n-1)+n^{2}+1\right)}{(d-1)^{2}}$ and $B^{(d)}_{n,d}= \frac{4Md\left[(3d-n-4)-d+1\right]}{(d-1)(d-2)}$. These results are valid for $d\neq1, \ 2$.

\subsection{Special case $d=2$}
\label{d2}
Although the main objective of this paper is to investigate solutions of the modified theories of gravity in a higher-dimensional spacetime, in this subsection we shall analyse the special case of planar $f(R)$ gravity with azimuthal symmetric spaces. The reason for this analysis resides in the fact that, in three-dimensions, the Einstein tensor, in a source-free space, is identically zero and only trivial solutions are possible. This fact happens because the Riemann tensor and the Ricci one have the same number of components. In fact, these tensors are related by \cite{Deser,Souradeep}:
\begin{equation}
\label{30}
R^{\mu\nu}_{\alpha\beta}=\epsilon^{\mu\nu\lambda}\epsilon_{\alpha\beta\gamma}\left(R^{\gamma}_{\lambda}-\frac{1}{2}\delta^{\gamma}_{\lambda}R\right) \ .
\end{equation} 
However, we would like to emphasize that in the modified theories of gravity, non-trivial vacuum solutions of the field equations arise. Considering $d=2$, the equations (\ref{Y1}) and (\ref{Y2}) can be written as 
\begin{equation}
\label{31}
2rF''-r\frac{Y'}{Y}F'-\frac{Y'}{Y}F=0
\end{equation}
and
\begin{equation}
\label{32}
ru''-\frac{1}{2}ru'\frac{Y'}{Y}+u'r\frac{F'}{F}-2u\frac{F'}{F}+u\frac{Y'}{Y}-u'=0 \ .
\end{equation}
By solving algebraically (\ref{31}) for $Y'/Y$ and substituting into (\ref{32}), we obtain a differential equation relating $F$ and $u$, namely  
\begin{equation}
\label{33}
u''-\left(\frac{rF''}{rF'+F}-\frac{F'}{F}+\frac{1}{r}\right)u'+\left(\frac{2F''}{rF'+F}-\frac{2F'}{rF}\right)u=0 \ ,
\end{equation}
whose general solution $u(r)$, for an arbitrary $F(r)$, is given by
\begin{equation}
\label{34}
u(r)=c_2r^{2}+c_1r^{2}\int \frac{e^{\int{\frac{r^{2}F''(r)F(r)-r^{2}F'(r)^{2}+F(r)^{2}}{\left(rF'(r)+F(r)\right)F(r)r}}dr}}{r^{4}}dr \ .
\end{equation}
\subsubsection{Solutions with constant curvature}
The field equations for constant curvature in the specific case of $d=2$, can be directly obtained from the general equations (\ref{eq1}) and (\ref{eq2}):
\begin{equation}
\label{35}
uv'+vu'=0
\end{equation} 
\begin{equation}
\label{36}
\frac{r}{2}\left(\frac{u'}{u}+\frac{v'}{v}\right)\left(\frac{r}{2}\frac{u'}{u}-1\right)-\frac{1}{2}\frac{r^{2}u''}{u}+\frac{ru'}{2u}=0 \ ,
\end{equation}
which provides the following solutions
\begin{equation}
\label{37}
v(r)=\frac{c_{0}}{u(r)}	\ \ {\rm and} \ \  u(r)=c_{0}+c_{1}r^{2}.
\end{equation} 
The scalar curvature associated with this metric, obtained from (\ref{Ricci}), is $R=-6c_{1}/c_{0}$. The scalar curvature, in (2+1) dimensions, assuming the existence of a cosmological constant $\Lambda$, is given by $R=6\Lambda$. In this case, we should choose $c_{1}=-\Lambda$. Furthermore, it is convenient, for reasons previously presented the choice of $c_{0}=1$. From this solution, we can notice that there is no Schwarzschild-type term in this (2+1) dimensional framework, for constant curvature. It was already expected because the different form of the Newtonian potential. 
  
\subsubsection{Solutions with $u(r)v(r)=$constant}
Also for this case it is easy to verify by using (\ref{31}) that $Y(r)=Y_0$, being $Y_0$ a constant, consequently $F(r)=Ar+B$. This form provides the following solution:
\begin{equation}
\label{38}
u(r)=c_{2}r^{2}+c_1\left[B(B-2Ar)+2A^{2}r^2\ln\left(A+B/r\right)\right] \ .
\end{equation}   
Apart from the factor $r^2$, the logarithmic term can be related with the Newtonian potential in two spatial dimensions, since in this case it has a logarithmic form. However, the details about the Newtonian potential is not our main goal, so it is interesting for our present purpose to find a way to avoid this logarithmic term. A natural way to remove it would be to consider $A=0$, but this is a trivial case which does not bring new feature to our analysis. An alternative procedure that we are going to adopt is to consider this analysis as an approximated one. This approach will be taken into account in the next section.
 
\subsubsection{An approximated solution}
Here we shall consider the form $F(r)=1+Ar$; however, implicitly assuming that $Ar<1$. This corresponds to a small perturbation on GR. In this approximation, all the quadratic terms in $Ar$ will be neglected. So, the solution given by (\ref{38}) will be written as
\begin{equation}
\label{39}
u(r)\approx c_{2}r^{2}+c_{1}-2Ac_{1}r \ .
\end{equation}
As we have already seen, one can always set $Y_{0}=1$, consequently $v(r)=1/u(r)$. We also consider $c_1=1$ and admitting the existence of a cosmological constant, it is natural to choose the integration constant $c_{2}=-\Lambda$. So we obtain:
\begin{equation}
\label{40}
u(r)\approx1-\Lambda r^{2}-2Ar \ .
\end{equation}  
The approximated result for the scalar curvature obtained by substituting (\ref{38}) into (\ref{Ricci}) is:
\begin{equation}
\label{41}
R\approx 6\Lambda+\frac{4A}{r}+12A^2\ln(r)+10A^2-24A^3r \ .
\end{equation}
The corresponding form of $f(R)$, in terms of the radial coordinate, $r$, is:
\begin{equation} 
\label{42}
f(r)\approx 4\Lambda+\frac{4A}{r}+8A^2(1+Ar)\ln(r)+\frac{20A^2}{3}-\frac{28A^3r}{3}+\frac{8A\Lambda r}{3} \ .
\end{equation}

\subsection{Solutions with $F(r)=F_0r^n$}
Another {\it Ansatz} for $F(r)$ that can also be examined is by considering $F(r)=F_0r^n$, with $n\neq 1$. For this case, equation (\ref{31}) provides the solution $Y(r)=Y_0r^{\frac{2n(n-1)}{n+1}}$, being $Y_0$ a constant. Substituting this result into (\ref{32}), we obtain for $u(r)$ the result below:
\begin{equation}
u(r)=c_1r^2+c_2r^{-\frac{2n}{n+1}}	\ .
\end{equation}
Considering, as in the previous analysis, the coefficient of the term squared in the radial distance associated with a "cosmological constant"', we may take $c_1=-\Lambda$. Finally for this case the scalar curvature and the corresponding function $f(R)$ read:
\begin{eqnarray}
R&=&-\frac{2(2n^2-5n-3)}{r^{\frac{2n(n-1)}{n+1}}}\frac\Lambda{Y_0(n+1)} \nonumber\\
f(R)&=&\frac{2F_0(n-1)}{n-3}\left[\frac{2(-2n^2+5n+3)\Lambda}{Y_0(n+1)}\right]^{\frac{n+1}{2(n-1)}}R^{\frac{n-3}{2(n-1)}} \ .
\end{eqnarray}

\subsection{The Chern-Simons action}
In three dimensional spacetime, it is possible to consider the Chern-Simons topological mass term to analyse the dynamics associated with gravitational field \cite{Deser,Souradeep}. In the context of $f(R)$ theories of gravity, the most general action in $(2+1)-$dimensions has the following form:
\begin{equation}
\label{CS}
S=\frac{1}{2\kappa}\int d^3x \sqrt{-g}f(R)+\frac{1}{\kappa\mu}{\cal S}_{cs}\ ,
\end{equation}
where
\begin{equation}
\label{CS1}
{\cal S}_{cs}=\frac{1}{4}\int d^2x \sqrt{-g}\epsilon^{\sigma\alpha\nu}\left[\omega_{\sigma}^{a}\partial_{\alpha}\omega_{\nu\alpha}+ \frac{1}{3}\omega_{\sigma}^{a}\omega_{\alpha}^{b}\omega_{\nu}^{c}\epsilon_{abc}\right] \ ,
\end{equation}
being $\mu$ an arbitrary constant with dimension of mass, and $\omega_{\sigma}^{a}$ the spin conection. The variation of $S$ with respect to the metric tensor provides the field equation,
\begin{equation}
\label{eqCS}
G_{\sigma\nu}+\frac{1}{\mu}C_{\sigma\nu}=T_{\sigma\nu}^{c} \ ,
\end{equation}
being $T_{\sigma\nu}^{c}$ the geometric energy-momentum tensor, and $C_{\sigma\nu}$ the Cotton-York tensor defined by
\begin{equation}
\label{CY}
C_{\sigma\nu}=\sqrt{-g}\ \epsilon_{\nu\alpha\beta}\nabla^{\alpha}\left(R^{\beta}_{\ \sigma}-\frac{1}{4}{\delta}^{\beta}_{\ \sigma}R\right) \ .
\end{equation}
For the azimuthal symmetric space under consideration, the only non-vanishing component of the Cotton-York tensor is:
\begin{eqnarray}
\label{compCS}
C_{02}&=&\frac{1}{8}\frac{\sqrt{Y}}{Y^4r}u\left[6u'''Y^2r^2+8u''Y^2r-8u'Y^2+4urY'^2+2uYY'+ 6u'Y'^2r^2\right.\nonumber\\&-&\left.2urYY''-9u''YY'r^2-8u'YY'r-3u'YY''r^2\right] \ ,
\end{eqnarray}
which must vanish in the case of conformally flat space. We can verify that de Sitter type solution (\ref{37}) satisfies the condition $C_{02}=0$, consequently is a solution of the field equation (\ref{eqCS}) with constant curvature.

\section{Conclusions}
\label{sec4}

In this paper we have analysed spherically symmetric vacuum solutions of $f(R)$ theories of gravity in $(1+d)-$dimensional spacetime. In order to achieve this objective, we have derived, by using the metric formalism, the field equations, which are expressed in terms of a general radial function $F(r)$. Considering different models, i.e., different {\it Ans\"atze} for this function, exact and approximated solutions have been found. All the exact cases considered generalize previous results obtained by Multamaki and Vilja in \cite{Mut}. As to the {\it Ansatz} $F(r)=1+Ar$, although we have obtained exact solutions for the field equations, we have found enormous difficulty in expressing $f(R)$ in terms of the scalar curvature and cosmological constant for $d\geq 4$. As an explicit example we have considered $d=4$. Moreover, we have also found exact solutions for the field equations considering $F(r)=F_0r$ and $F(r)=F_0r^n$. For these cases the corresponding expression for $f(R)$ have been obtained. A further case that we have also studied is for $F(r)=1+Ar^n$; unfortunately this case can only be handled approximately; for this case we have given the solutions for $u(r)$ and corresponding scalar curvature $R$.

Although the main objective of this paper is the analysis of modified theories of gravity in higher dimensions, we also have focused our attention to the case $d=2$. Exact solutions for different {\it Ans\"atze} have been found. For $F(r)=Ar+B$, a specific approach, namely considering $Ar<1$, can avoid logarithmic terms arising in the solution for the metric tensor components. These logarithmic terms, on the other hand, could be associated with Newtonian potential which we have decided, for simplicity only, do not deal with. For $F(r)=F_0r^n$, with $n\neq 1$, we have explicitly exhibited exact solutions for $u(r)$, which presents a singular behavior for positive value of $n$. Finally, we have briefly considered the inclusion of the Chern-Simons topological term for gravity. In this context we have shown that de Sitter solution is also solution of the complete equation of motion.

Many researcher have studied different $f(R)$ models recently. The purpose of this paper is to provide a general approach about how to obtain exact and approximated solutions of the field equation considering an arbitrary radial function $F(r)$ in a time independent spherically symmetric higher-dimensional spacetime. Although we have consider only vacuum solutions, we have reasons to believe that many subjects still deserve to be investigated in higher-dimensional spacetime scenario. A natural extension could be to consider the presence of matter fields, to derive the field equations and develop a similar analysis as we have done here. 

\section{Acknowledgment}

TRPC thanks CAPES for financial support. ERBM thanks Conselho Nacional de Desenvolvimento Cient\'\i fico e Tecnol\'ogico (CNPq.)  for partial financial support, FAPESQ-PB/CNPq. (PRONEX) and FAPES-ES/CNPq. (PRONEX).

\appendix
\section{Explicit derivation of the solution for $Y(r)=$constant}\label{A}
Here in this Appendix we briefly present some steps adopted in the obtainment of (\ref{14}), solution of the field equations (\ref{Y1}) and (\ref{Y2}) for $Y(r)=Y_0=$constant. First, let us  focus our attention on the corresponding homogeneous equation of (\ref{Y2}). Defining a new variable $z=\frac{Ar}{B}$, we have:
\begin{equation}
\label{47}
z^2(z+1)u''(z)+\left[z^2(d-2)+z(d-3)\right]u'(z)-2\left[z(d-1)+d-2\right]u(z)=0 \ ,
\end{equation}
whose solution is 
\begin{equation}
\label{48}
u_{h}(z)=az^2+bz^{2-d}\left[d\Phi\left(-\frac{1}{z},1,d\right)-1\right]\ ,
\end{equation}
where $a$ and $b$ are integration constants. The above solution is expressed in terms of  the Phi-function \cite{grad}, defined as shown below:
\begin{equation}
\label{49}
\Phi\left(v,s,\alpha\right)=\sum^{\infty}_{n=0}\frac{v^n}{(\alpha+n)^s}\ ,
\end{equation}
which converges for $\left|v\right|<1$ and $\alpha\neq 0, \ -1, \ -2, \ ... $ \ . It is simple to verify the following identity, for $z>1$:
\begin{equation}
\label{50}
\left(-\frac{1}{z}\right)^d\left[\Phi\left(-\frac{1}{z},1,d\right)-1\right]=-d\left[\sum^{d}_{n=1}\frac{1}{n(-z)^n}+\ln\left(1+\frac{1}{z}\right)\right].
\end{equation}
Substituting the above expression into (\ref{48}) and rescaling the integration constants, the solution can rewritten as
\begin{equation}
\label{51}
u_{h}(z)=az^2+bz^2\left[\ln{\left(1+\frac{1}{z}\right)}+\sum^{d}_{n=1}\frac{1}{n(-z)^n}\right].
\end{equation}
Finally, by recovering the original radial coordinate, $r$, we can express this solution as follows:
\begin{equation}
\label{52}
u_{h}(r)=c_2r^2+c_1r^2\left[\ln{\left(1+\frac{B}{Ar}\right)}+\sum^{d}_{n=1}\frac{1}{n(-\frac{Ar}{B})^n}\right].
\end{equation}  

The particular solution, $u_p(r)$, of the field equations is straightforwardly obtained. So, taking into account both solutions, we can write $u(r)=u_p(r)+u_h(r)$, which reads 
\begin{eqnarray}
\label{55}
u(r)&=&c_2r^2+c_1r^2\left[\ln{\left(1+\frac{B}{Ar}\right)}+\sum^{d}_{n=1}\frac{(-1)^n}{nr^n}\left(\frac{B}{A}\right)^n\right]+Y_0\left[\frac{2}{d-1}\frac{A^2r^2}{B^2}\ln\left(1+\frac{B}{Ar}\right)\right.\nonumber\\
&+&\left.1-\frac{2}{d-1}\frac{Ar}{B}\right] \ . 
\end{eqnarray}

\end{document}